\newcommand{\be}{\begin{equation}}
\newcommand{\ee}{\end{equation}}
\newcommand{\bea}{\begin{eqnarray}}
\newcommand{\eea}{\end{eqnarray}}
\newcommand{\vecvar}[1]{\mbox{\boldmath$#1$}}
\newcommand{\ri}{{\rm i}}
\newcommand{\rd}{{\rm d}}
\newcommand{\re}{{\rm e}}
\newcommand{\ket}[1]{\left|#1\right\rangle}
\newcommand{\bra}[1]{\left\langle#1\right|}
\newcommand{\braket}[2]{\left\langle#1|#2\right\rangle}

\newcommand{\eref}[1]{Eq.~(\ref{#1})}

\newcommand{\fref}[1]{Fig.~\ref{#1}}
\newcommand{\Fref}[1]{Figure \ref{#1}}

\documentclass[preprint]{jpsj2}
\title{Temperature dependence of ESR intensity for the nanoscale 
molecular magnet ${\rm V}_{15}$}

\author{
Manabu \textsc{Machida}$^{1}$
\thanks{E-mail address: machida@iis.u-tokyo.ac.jp}, 
Toshiaki \textsc{Iitaka}$^{2}$
\thanks{E-mail address: tiitaka@riken.jp}, and 
Seiji \textsc{Miyashita}$^{3}$
\thanks{E-mail address: miya@spin.phys.s.u-tokyo.ac.jp}
}

\inst{
$^{1}$Institute of Industrial Science, The University of Tokyo, 
4-6-1 Komaba, Meguro-ku, Tokyo 153-8505, Japan\\
$^{2}$Computational Astrophysics Laboratory, RIKEN 
(The Institute of Physical and Chemical Research), 2-1 Hirosawa, Wako, 
Saitama 351-0198, Japan\\
$^{3}$Department of Physics, Graduate School of Science, 
The University of Tokyo, 7-3-1 Hongo, Bunkyo-ku, Tokyo 113-0033, Japan
}

\recdate{\today}

\abst{
The electron spin resonance (ESR) of nanoscale molecular magnet 
${\rm V}_{15}$ is studied.  
Since the Hamiltonian of ${\rm V}_{15}$ has a large Hilbert space and 
numerical calculations of the ESR signal evaluating the Kubo formula 
with exact diagonalization method is difficult, 
we implement the formula with the help of the random vector technique 
and the Chebyshev polynominal expansion, 
which we name the double Chebyshev expansion method.  
We calculate the temperature dependence of the ESR intensity of 
${\rm V}_{15}$ and compare it with the data obtained in experiment.  
As another complementary approach, we also implement the Kubo formula with 
the subspace iteration method taking only important low-lying states 
into account.  
We study the ESR absorption curve below $100{\rm K}$ by means of 
both methods.  
We find that side peaks appear due to the Dzyaloshinsky-Moriya 
interaction and these peaks grows as temperature decreases.   
}

\kword{electron spin resonance, molecular magnets}

\begin{document}
\maketitle

The ${\rm V}_{15}$ molecule is the complex of formula 
${\rm K}_6\left[{\rm V}_{15}^{\rm IV}{\rm As}_6{\rm O}_{42}
\left({\rm H}_2{\rm O}\right)\right]\cdot 8{\rm H}_2{\rm O}$.  
Since M\"{u}ller and D\"{o}ring synthesized this molecule 
for the first time,
\cite{Muller88a,Muller91a,Gatteschi91a,Barra92a,Gatteschi93a} 
${\rm V}_{15}$ has been studied intensively as one of 
promising nanometer-scale molecular magnets.  
In ${\rm V}_{15}$, fifteen $1/2$ spins of vanadium ions are arranged 
almost on a sphere.  The triangle cluster with three spins is sandwiched 
by the upper and lower hexagons.  
In experiment, an adiabatic change of the magnetization has been 
observed in a fast sweeping field.\cite{Chiorescu00a,Chiorescu00b,Chiorescu03a}
In a slow sweeping field, an interesting magnetic plateau appears.
This phenomenon, called the phonon bottleneck effect, is due to
an effect of contact with thermal bath,\cite{Chiorescu00a,Chiorescu00b} 
and also theoretically analyzed from a general viewpoint of 
the magnetic Foehn effect.\cite{MFE}
Since the magnetization changes smoothly at zero field from $-1/2$ to $1/2$ 
when the field is swept adiabatically at low temperatures, 
the gap between the ground state and the lowest excited state 
at zero field is believed to open. The most plausible origin of
the gap is the Dzyaloshinsky-Moriya (DM) interaction.
\cite{Dzyaloshinsky58a,Moriya60a,Moriya60b,MN,Chiorescu03a,Konstantinidis02a}
Around $2.8{\rm T}$, the ground state magnetization changes
from $1/2$ to $3/2$. 
This change also occurs smoothly, and this broadness of the change is 
also considered to be caused by the DM interaction.\cite{Chiorescu00b}  
However, the detail of the mechanism of this broad change is not fully 
understood yet.\cite{DeRaedt04a,DeRaedt04b}   

One possible way to determine the mechanism of adiabatic change in 
${\rm V}_{15}$ is that we compare the numerically obtained ESR 
absorption curve of ${\rm V}_{15}$ with the experimentally obtained one.  
With this motivation, in this paper, we establish numerical methods 
to study the ESR absorption curve obtained 
from the model Hamiltonian of ${\rm V}_{15}$.  
The ESR absorption curve is calculated by the Kubo formula 
in theory.\cite{Kubo54a,Kubo57a}  The direct implementation of the Kubo 
formula is to diagonalize the Hamiltonian matrix.\cite{MYO} 
However, this direct method is impossible in most cases 
because the dimension of the Hilbert space is quite large.  
For example, it is $2^{15}=32768$ for ${\rm V}_{15}$.  

In this paper, we study the ESR absorption curve of ${\rm V}_{15}$ by 
proposing two methods: one is {\it the double Chebyshev expansion method} 
(DCEM) and the other is {\it the subspace iteration method} (SIM).  
Relying on these methods, we reveal the temperature dependence of 
the intensity of ${\rm V}_{15}$, and find side peaks due to 
the DM interaction which allows transitions between excited states 
otherwise forbidden.  

\begin{figure}[tb]
\begin{center}
\hspace{-10mm}
\includegraphics[scale=0.4,clip]{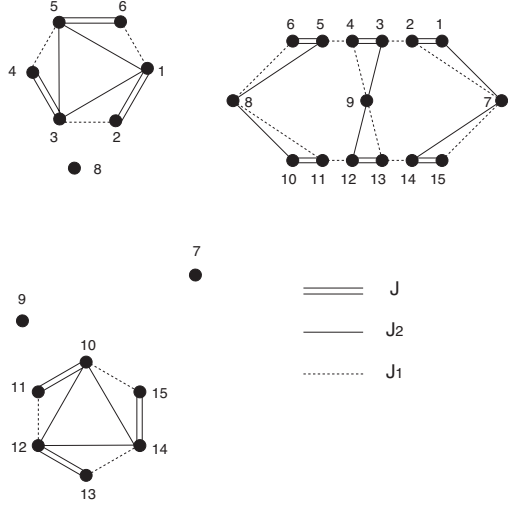}
\end{center}
\caption{The figures on the left show interactions within 
each hexagon.  The interaction within the triangle is negligibly small.  
The figure on the top right shows interactions between the triangle 
and the upper and lower hexagons.  
}
\label{f0}
\end{figure}

\begin{figure}[tb]
\begin{center}
\includegraphics[scale=1.2,clip]{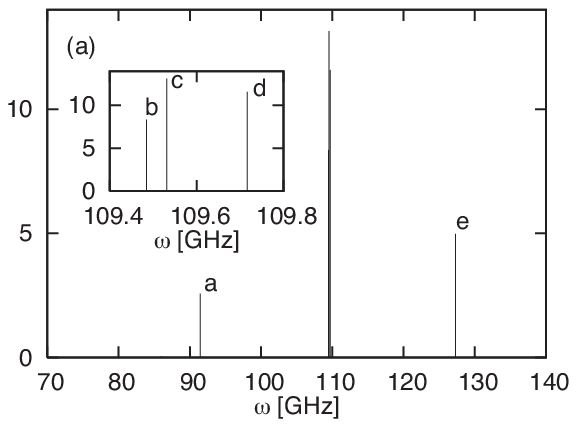}
\includegraphics[scale=1.2,clip]{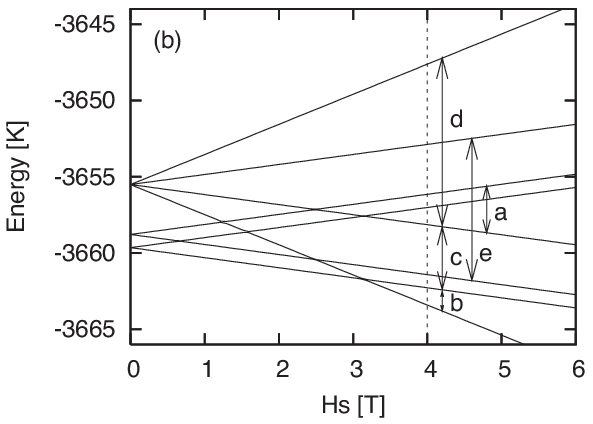}
\end{center}
\caption{
(a) The ESR absorptions calculated by the SIM at $4{\rm T}$ and 
$32{\rm K}$ are shown.  The height of 
the pulse at each resonant frequency expresses the value of 
the coefficient of the delta function of the absorption.  
The inset shows the magnified figure around $110 {\rm GHz}$.  
(b) The lowest eight energy levels of ${\rm V}_{15}$.  
The vertical dashed line at $H_{\rm S}=4{\rm T}$ shows the magnetic 
field applied to ${\rm V}_{15}$.  The arrows denote 
transitions corresponding peaks labeled by a through e in (a).  
}
\label{f3}
\end{figure}

The magnetic interactions in ${\rm V}_{15}$ are described by 
the following Hamiltonian\cite{DeRaedt04a,DeRaedt04b}  
\be
\mathcal{H} = 
- \sum_{\langle i,j\rangle}J_{ij}\vecvar{S}_i\cdot\vecvar{S}_j 
+ \sum_{\langle i,j\rangle}\vecvar{D}_{ij}\cdot
\left[\vecvar{S}_i\times\vecvar{S}_j\right]
- \sum_i\vecvar{H}_{\rm S}\cdot\vecvar{S}_i.
\label{hami}
\ee
We show the interactions between spins in \fref{f0}.  
For $J_{ij}$, we have three different values $J$, $J_1$, and 
$J_2$ ($|J|>|J_2|>|J_1|$) with 
respect to the bonds on the upper and lower hexagons.  Three spins on 
the triangle between two hexagons interact with the hexagons by $J_1$ 
and $J_2$.  The interactions between the three spins are 
negligibly small.  
Here we take 
$J=-800{\rm K}$, $J_2=-350{\rm K}$, and $J_1=-225{\rm K}$.
\cite{Konstantinidis02a}.  
The second term on the right-hand side in \eref{hami} describes the 
DM interaction.  DM vectors are considered to exist on the two hexagons 
in the bonds with $J$.  
We take the reference DM vector 
$\vecvar{D}_{1,2}$ to be 
$D_{1,2}^x=D_{1,2}^y=D_{1,2}^z=40{\rm K}$.\cite{DeRaedt04a}  
The other DM vectors on the upper hexagon are obtained by rotating 
$\vecvar{D}_{1,2}$ by $2\pi/3$ and $4\pi/3$, i.e., 
$D_{3,4}^x=14.641{\rm K}$, $D_{3,4}^y=-54.641{\rm K}$, 
$D_{3,4}^z=40{\rm K}$, 
$D_{5,6}^x=-54.641{\rm K}$, $D_{5,6}^y=14.641{\rm K}$, and 
$D_{5,6}^z=40{\rm K}$.  
If we assume the $D_3$ symmetry of ${\rm V}_{15}$,\cite{Muller88a} 
the lower hexagon differs from the upper hexagon by rotation $\pi/6$, 
and the $z$ components of the DM vectors on the lower hexagon have 
opposite sign from those of the DM vectors on the upper hexagon.  
Thus, the DM vectors on the lower hexagon are obtained as 
$D_{10,11}^x=-54.641{\rm K}$, $D_{10,11}^y=-14.641{\rm K}$, 
$D_{10,11}^z=-40{\rm K}$, 
$D_{12,13}^x=40{\rm K}$, $D_{12,13}^y=-40{\rm K}$, $D_{12,13}^z=-40{\rm K}$, 
$D_{14,15}^x=14.641{\rm K}$, $D_{14,15}^y=54.641{\rm K}$, and 
$D_{14,15}^z=-40{\rm K}$.  
We assume the magnetic field is applied parallel to the $c$-axis of 
the molecule ($z$-axis) and set 
$\vecvar{H}_{\rm S}=(0,0,H_{\rm S})$.  We take $H_{\rm S}=4{\rm T}$ 
throughout the paper (see also \fref{f3}(b)).  

\textit{Double Chebyshev expansion method (DCEM)}. --- 
This method realizes $O(N)$ calculation with the help of the random 
vector technique\cite{DeRaedt00a,Iitaka04a} and 
the Chebyshev polynomial expansion of exponential operators.  
The DCEM is an extension of the 
Boltzmann-weighted time-dependent method (BWTDM) developed by 
Iitaka and Ebisuzaki.\cite{Iitaka03a}  
We briefly review the BWTDM.  The procedure of the BWTDM is 
divided by the following five steps.  
By adopting the unit in Ref.~22, we have dimensionless energy, time, 
and temperature.  
At the first step, we prepare a random vector 
$\ket{\Phi}$.  For a given basis $\ket{n}$ of the Hilbert space, this 
random vector is given by $\ket{\Phi}=\sum_{n=1}^N\ket{n}\xi_n$.  Here, 
the dimension of the Hilbert space is $N$ and the statistical average 
of the complex random coefficients $\{\xi_n\}$ satisfies 
$\langle\langle\xi_{n'}^*\xi_n\rangle\rangle=\delta_{n'n}$.   
At the second step, we obtain the Boltzmann-weighted random vector 
$\ket{\Phi_{\rm Boltz}}=\re^{-\beta\mathcal{H}/2}\ket{\Phi}$ 
making use of the Chebyshev polynomial expansion.  Here, 
$\beta (=1/T)$ is the inverse temperature.  To this end, 
we divide the Hamiltonian by constant $\Delta\lambda$ as 
$\mathcal{H}_{\rm sc}=\mathcal{H}/\Delta\lambda$ 
so that the largest eigenvalue of $\mathcal{H}_{\rm sc}$ does not exceed 
unity.  Then, we expand $\re^{-\beta\mathcal{H}/2}$ as follows.  
\bea
\re^{-\beta\mathcal{H}/2} &=& 
I_0\left(-\beta\Delta\lambda/2\right)T_0(\mathcal{H}_{\rm sc}) 
\nonumber \\
&+& 2\sum_{k=1}^{k_{\max}}
I_k\left(-\beta\Delta\lambda/2\right)T_k(\mathcal{H}_{\rm sc}),
\eea
where $I_k(x)$ is the modified Bessel function and 
$T_k(\mathcal{H}_{\rm sc})$ is the Chebyshev polynomial, which satisfies 
$T_k(\mathcal{H}_{\rm sc})=
2\mathcal{H}_{\rm sc}T_{k-1}(\mathcal{H}_{\rm sc})-
T_{k-2}(\mathcal{H}_{\rm sc})$, 
$T_0(\mathcal{H}_{\rm sc})=1$, and 
$T_1(\mathcal{H}_{\rm sc})=\mathcal{H}_{\rm sc}$.  
By this procedure, we obtained $\re^{-\beta\mathcal{H}/2}$ 
without diagonalization.  
At the third step, we obtain time evolutions of vectors 
$\ket{\Phi_{\rm Boltz};t}=\re^{-\ri\mathcal{H}t}\ket{\Phi_{\rm Boltz}}$ 
and 
$\ket{\Phi_{M^x};t}=\re^{-\ri\mathcal{H}t}\ket{\Phi_{M^x}}$, where 
$\ket{\Phi_{M^x}}=M^x\ket{\Phi_{\rm Boltz}}$ and 
$M^x=\sum_{j=1}^{N_{\rm spin}}S^x_j$.  In the BWTDM, 
the time evolution is performed by the leap frog method
\cite{Iitaka94a}, which evolves state $\ket{\phi;t}$ as 
\be
\ket{\phi;t+\Delta t}=
-2\ri\mathcal{H}\Delta t\ket{\phi;t}+\ket{\phi;t-\Delta t}.  
\ee
Note that the condition $E_{\rm max}\Delta t\ll 1$ should be satisfied, 
where $E_{\rm max}$ is the largest eigenvalue of the Hamiltonian.  
At the fourth step, we calculate the correlation function 
\bea
g(t;T) &=& \langle M^xM^x(t)\rangle \nonumber \\
&=& 
{\rm Tr}\left[\re^{-\beta\mathcal{H}}M^x\re^{\ri\mathcal{H}t}M^x
\re^{-\ri\mathcal{H}t}\right]/
{\rm Tr}\left[\re^{-\beta\mathcal{H}}\right] \nonumber \\
&=& \frac{\langle\langle\;\bra{\Phi_{M^x};t}M^x\ket{\Phi_{\rm Boltz};t}\;
\rangle\rangle}{
\langle\langle\;\braket{\Phi_{\rm Boltz}}{\Phi_{\rm Boltz}}\;
\rangle\rangle},
\eea
where the trace was replaced by the inner product of random vectors.
Finally, the imaginary part of the dynamical susceptibility 
$\chi''(\omega;T)$ is obtained by the Fourier transform of 
$g(t;T)$.  
\bea
\chi''(\omega;T) &=& \left(1-\re^{-\beta\omega}\right){\rm Re}
\int_0^{\infty} g(t;T) \re^{-\ri\omega t} \rd t \nonumber \\
&\hspace{-30mm}=& \hspace{-15mm}\left(1-\re^{-\beta\omega}\right){\rm Re}
\int_0^{T_{\rm max}} g(t;T) \re^{-\ri\omega t} \re^{-\eta^2t^2/2} \rd t.
\eea
Here, we introduced the Gaussian filter with variance $1/\eta^2$.  
This $\eta$ determines the frequency resolution.  The upper limit of the 
integral $T_{\rm max}$ satisfies 
$T_{\rm max}\sim 1/\eta$ in order to avoid the Gibbs oscillation.  
Also $\eta$ should satisfy 
$0<\eta\ll 1$, $\eta\ll H_{\rm S}$, and $\beta\eta^2\ll H_{\rm S}$.  
The average energy absorption per unit time $I(\omega;T)$ 
is given by 
\be
I(\omega;T)=\frac{\omega H_{\rm R}^2}{2}\chi''(\omega;T).
\ee

The DCEM is almost the same as the BWTDM.  Only the third step is 
different.  In the DCEM, we make use of the Chebyshev polynomial 
expansion not only in the second step obtaining $\re^{-\beta\mathcal{H}/2}$, 
but also in the third step of time evolution.  State vector 
$\ket{\phi;t}$ is evolved with the Chebyshev polynomials as 
\bea
\ket{\phi;t+\tau} &=& 
\re^{-\ri\tau\Delta\lambda\mathcal{H}_{\rm sc}}\ket{\phi;t} \nonumber \\
&=& J_0(\tau\Delta\lambda)T_0(\mathcal{H}_{\rm sc})\ket{\phi;t} 
\nonumber \\
&+& 2\sum_{k=1}^{k_{\rm max}}(-\ri)^k
J_k(\tau\Delta\lambda)T_k(\mathcal{H}_{\rm sc})\ket{\phi;t},
\eea
where $J_k(x)$ is the Bessel function.  Note that time step $\tau$ is 
not necessarily small.  
In the ESR experiment for ${\rm V}_{15}$, magnetic field $H_{\rm S}$ 
$(\sim 1 {\rm K})$ is usually much smaller than the strongest coupling 
$J_{\rm max}(=|J|)$ between spins $(\sim 10^3 {\rm K})$.  That is why 
the frequency 
of precession of the spins is rather small.   This means that 
we need to evolve state vectors long time but do not need fine 
resolution of time step in order to detect small frequencies of 
spin precession.  Since it is possible for time step $\tau$ to have 
larger value than time step $\Delta t$, the DCEM is more efficient than 
the BWTDM for low magnetic fields.  
Table \ref{t1} shows typical computation times with the DCEM and 
the BWTDM.  Using asymptotic behavior of the Bessel function, 
the ratio of two computation times is evaluated as
\be 
\frac{\left[\mbox{Chebyshev}\right]}{\left[\mbox{Leap-frog}\right]}
\sim \frac{a H_{\rm S}}{J_{\rm max}}
\ln\left[\frac{b}{H_{\rm S}/J_{\rm max}}\right], 
\ee
where $a$ and $b$ are constants.  

\begin{table}[t]
\caption{Typical computation times of two methods at 
various magnetic fields.}
\label{t1}  
\begin{tabular}{crrr}
\hline
    & 1000(T) & 100(T) & 10(T) \\
\hline
Chebyshev & 126(min) & 187(min) & 430(min) \\
\hline
Leap-frog & 11(min) & 165(min) & 1326(min) \\
\hline
\end{tabular}
\end{table}

\begin{figure}[tb]
\begin{center}
\hspace{-10mm}
\includegraphics[scale=1.2,clip]{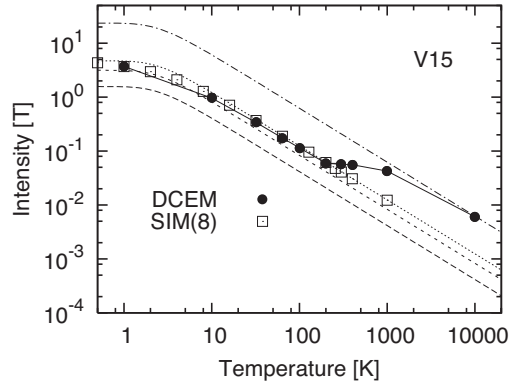}
\end{center}
\caption{
The temperature dependence of the intensity of ${\rm V}_{15}$ is shown.  
Solid circles denote data by the DCEM and open squares denote data by 
the SIM with the lowest eight levels.  
Dashed lines denote, from the bottom, intensities of independent 
1, 2, 3, and 15 spins, respectively.  
}
\label{f1}
\end{figure}

We first study the temperature dependence of the intensity 
$I(T)\;\left(=\int_0^{\infty}I(\omega;T) \rd\omega\right)$ 
using the DCEM.  
In \fref{f1}, we show the temperature dependence of $I(T)$.  
The intensity $I(T)$ obtained by the DCEM is denoted by solid circles.  
The open squares in \fref{f1} denote data obtained by the subspace 
iteration method, which we will explain below.  
At very high temperatures, each spin in ${\rm V}_{15}$ acts 
as an isolated spin.  Therefore, the intensity should be given by 
the multiplication of $I_1(T)$ by 15, where 
$I_1(T)=\frac{\pi}{8}H_{\rm R}^2H_{\rm S}\tanh\left(
\beta H_{\rm S}/2\right)$ 
is the intensity of an isolated spin.  
In \fref{f1}, $15\times I_1(T)$, $3\times I_1(T)$, 
$2\times I_1(T)$, and $1\times I_1(T)$ are also shown by 
dash-dotted line, dotted line, short-dashed line, and dashed line, 
respectively.  As temperature decreases, the effective 
number of spins changes from 15 to 3.  Note that the ground state 
magnetization is $3/2$ in the present case of $H_{\rm S}=4 {\rm T}$ 
(see also the energy diagram in \fref{f3}(b)).  
In experiment, the intensity starts deviating the curve for three spins at 
around $200 {\rm K}$.\cite{Ajiro03a}  Here we confirm that the intensity 
actually follows the curve for independent 15 spins at high 
temperatures.  
 
\textit{Subspace iteration method (SIM)}. --- 
We explain another method obtaining the ESR absorption curve 
of large molecules.  At low temperatures, the ESR absorption 
occurs only from transitions among low-lying energy levels because 
states of the system are confined near the ground state.  
The SIM is a numerical method that 
takes a small subspace of the total Hilbert space which concerns 
only low-lying states.\cite{Chatelin88a,Mitsutake96a}  
Within this subspace, the ESR absorption curve 
is obtained by the explicit formulation of the Kubo formula.  That is, 
by direct diagonalization of the small Hamiltonian.  
We obtain the eigenvalues and eigenvectors of the subspace and 
obtain the imaginary part of the susceptibility $\chi''(\omega)$;
\cite{MYO} 
\bea
\chi''(\omega) &=& \frac{\pi}{Z}\sum_{m,n}
\left(\re^{-\beta E_m}-\re^{-\beta E_n}\right)
|\bra{\psi_m}M^x\ket{\psi_n}|^2 \nonumber \\
&& \times\delta(\omega-(E_n-E_m)),
\eea   
where $Z$ is the partition function.  

The calculation in the DCEM is numerically exact, but it generally requires 
longer computation time than that of the SIM.  On the other hand, 
the calculation in the SIM is carried out shorter time for a moderate 
value of $\tilde{N}$, which is enough to study the property at low 
temperatures.  However, it ignores higher states completely.  
The two methods are complementary to each other.  

In \fref{f1}, the intensities obtained by the SIM with the lowest eight levels 
$(\tilde{N}=8)$ are plotted by open squares.  Those are consistent with 
the solid circles obtained by the DCEM at temperatures lower than 
$200{\rm K}$.  
Although the data of the DCEM start deviating from the line of three spins 
at around $200{\rm K}$, the data of the SIM stay at the line, which reflects 
the fact that the SIM considers only the lowest eight levels.  
Here we confirm that the SIM can reproduce the data of the DCEM at low 
temperatures with a shorter computation time.  Thus, in \fref{f1}, 
we can have more points at low temperatures.  

We point out the following two points, which will be discussed in more 
detail elsewhere.\cite{Machida05a}  
First, the data slightly deviate from the dotted line at low 
temperatures as seen in the experiment by Sakon et al.\cite{Sakon04a}  
For the present field $(H_{\rm S}=4{\rm T})$, i.e., 
the ground state magnetization is $3/2$, the ratio $I(T)/I_1(T)$ takes 
values smaller than 3 at low temperatures due to the DM interaction.  
Second, in the experiment by Ajiro, et al.,\cite{Ajiro03a} 
the intensity follows the line of an isolated $1/2$ spin ($I(T)/I_1(T)=1$).  
This is because the ground state magnetization is $1/2$ in the field 
weaker than $2.8 {\rm T}$.  

\begin{figure}[tbh]
\begin{center}
\includegraphics[scale=1.2,clip]{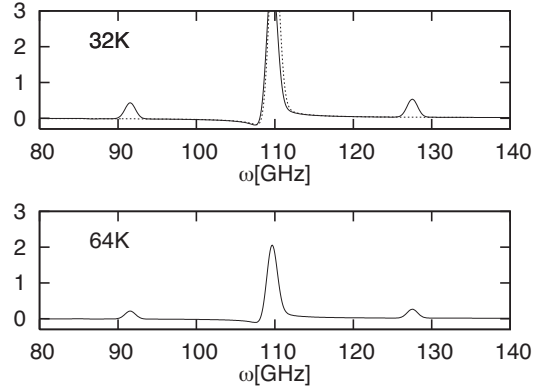}
\end{center}
\caption{
The upper and lower panels show the ESR absorption curves of ${\rm V}_{15}$ 
at $4{\rm T}$ and at $32{\rm K}$ and $64{\rm K}$, respectively.  
In the upper panel, the dashed curve shows the absorption curve 
obtained from the Hamiltonian without the DM interaction.  
}
\label{f2}
\end{figure}

Finally, let us study the absorption curve $I(\omega;T)$.  
\Fref{f2} compares $I(\omega;T=32{\rm K})$ and $I(\omega;T=64{\rm K})$ 
obtained by the DCEM.  
Note that the widths of the peaks are solely due to the finite frequency 
resolution of the DCEM given by $\eta$.  We see that the peaks grow 
as temperature decreases. 
In the upper panel in \fref{f2}, intensities with and without the DM 
interaction are compared.  We conclude that the 
side peaks near the main peak around $110 {\rm GHz}(\simeq 4{\rm T})$ appear 
due to the DM interaction.  
   
Figures \ref{f3}(a) and \ref{f3}(b) show results obtained by the SIM.  
In \fref{f3}(a), the ESR absorptions by the lowest eight levels 
($\tilde{N}=8$) are calculated at $32 {\rm K}$.  The height of 
pulses shows the value of coefficients of delta functions of 
the absorptions.  The peak around $110 {\rm GHz}$ consists of three 
resonant peaks, whereas each side peak consists of single resonant peak. 
\Fref{f3}(b) shows the lowest eight energy levels of ${\rm V}_{15}$ 
as a function of field $H_{\rm S}$.  
Each labeled peak in \fref{f3}(a) comes from the corresponding labeled 
transition between energy levels shown in \fref{f3}(b).  

Thus, the combination of the DCEM and the SIM provides a powerful method 
to elucidate the ESR of large molecular magnets 
such as ${\rm V}_{15}$.  

This work is supported by the Grant-in-Aid from the Ministry of 
Education, Culture, Sports, Science and Technology, and also by 
NAREGI Nanoscience Project, Ministry of Education, Culture, Sports, 
Science and Technology, Japan. The simulations were partially 
carried out by using the computational facilities of the Super 
Computer Center of Institute for Solid State Physics, the University 
of Tokyo, and Advanced Center for Computing and Communication, 
RIKEN (The Institute of Physical and Chemical Research).

\end{document}